\documentclass[aps,prl,preprint,superscriptaddress,floatfix,altaffilletter]{revtex4}

\usepackage{graphicx}
\usepackage{dcolumn}
\usepackage{bm}
\usepackage{amsmath}
\usepackage{multirow}
\usepackage{hhline}

\begin{document}

\title{Thermodynamic limits to energy conversion in solar thermal fuels}

\date{\today}
\author{David A Strubbe \footnote{dstrubbe@ucmerced.edu}}
\affiliation{Department of Materials Science and Engineering, Massachusetts Institute of Technology, Cambridge, MA 02139}
\affiliation{Department of Physics, University of California, Merced, CA 95348}
\author{Jeffrey C Grossman}
\affiliation{Department of Materials Science and Engineering, Massachusetts Institute of Technology, Cambridge, MA 02139}

\begin{abstract}

Solar thermal fuels (STFs) are an unconventional paradigm for solar energy conversion and storage which is attracting renewed attention.
In this concept, a material absorbs sunlight and stores the energy chemically via an induced structural change, which can later be reversed
to release the energy as heat. An example is the azobenzene molecule which has a {\it cis}-{\it trans} photoisomerization with these properties,
and can be tuned by chemical substitution and attachment to templates such as carbon nanotubes, small molecules, or polymers.
By analogy to the Shockley-Queisser limit for photovoltaics, we analyze the maximum attainable efficiency for STFs from fundamental thermodynamic considerations.
Microscopic reversibility provides a bound on the quantum yield of photoisomerization due to fluorescence, regardless of details of photochemistry.
We emphasize the importance of analyzing the free energy, not just enthalpy, of
the metastable molecules, and find an efficiency limit for conversion to stored chemical energy equal to the Shockley-Queisser limit.
STF candidates from a recent high-throughput search are analyzed in light of the efficiency limit.


\end{abstract}

\maketitle

  Solar thermal fuels (STFs) are an unconventional paradigm for solar-energy harvesting and storage, which provides long-term
  storage as chemical energy and later release as heat.
  Unlike in photovoltaics (PV), incident solar photons are not converted to electricity but rather drive a reversible structural change in
  a material. Molecules that undergo a structural change on absorption of light (photoisomerization) are referred to as ``photochromic,''
  since in general the optical absorption spectrum will change with the new structure. (STFs have also been referred to as ``molecular
  solar thermal'' (MOST) \cite{poulsen_EES}.)
  Various classes of photochromic molecules are known, such as azobenzene, spiropyran/merocyanine,
  norbornadiene/quadricyclane, and fulvalene(tetracarbonyl)diruthenium \cite{Kucharski_review}. The basic concept was developed decades ago \cite{Bolton},
  but available molecular materials did not have adequate performance to enable applications, with regard to metrics such as
  cyclability, stored energy density, visible light absorption, and cost. Modern advances in nanoscience and atomistic computation and design
  have given new approaches and interest in this idea, as molecular and nanoscale templates and functionalization have produced increases
  in stored energy density and lifetime \cite{Kolpak_NanoLett, Kolpak_JCP, Kucharski_NatChem, Liu, Durgun, Han, Quant, Feng}, and performance of solution-based
  \cite{poulsen_EES, Wang} and solid-state devices \cite{Zhitomirsky} have been demonstrated. While current devices deliver stored energy as
  heat, it may also be possible to use photo-induced mechanical motion \cite{Bardeen} to convert the stored energy to other forms \cite{Salzbrenner}.

  The question of the actual efficiency of STF devices, taking together all the relevant material properties, is a crucial one for assessing
  the relevance of STFs as an approach for solar-energy conversion, especially by comparison to the more established PV, solar thermal, and
  solar fuels technologies.
  However, the efficiency has remained unclear: it has been estimated experimentally in only a few cases \cite{poulsen_EES},
  and given only preliminary and somewhat limited theoretical analysis in the literature \cite{Bren, Kucharski_NatChem, Borjesson}.
  These works have focused primarily on
  enthalpy but not considered free energy or the key roles of chemical equilibrium, entropy, and temperature, and have relied on idealized
  or arbitrary parameters for simplicity. Other work has analyzed the photochemistry in detail but not overall device efficiency \cite{Bolton, Almgren},
  or considered schemes more general than STFs \cite{Ross}.

  In the field of PV, the well-known work of Shockley and Queisser \cite{shockley} (hereafter, SQ) bridged the gap between analysis of the
  specific PV materials, and analysis of general heat engines, to find an efficiency limit for the single-junction PV scheme under sunlight,
  with constraints not from the properties of current materials but from rigorous thermodynamics. They found that the maximum efficiency
  attainable for a single-junction cell at 300 K in unconcentrated sunlight is 32\%, achieved for a bandgap of 1.27 eV.

  In this paper, we follow the SQ analysis
  to derive formulae for the efficiency of STFs and their limits from rigorous thermodynamic considerations. We underscore the detailed analogy to PV,
  including $I$-$V$ characteristics, despite the differing device operation; show the importance of the free energy;
  find a limit to the quantum yield of photoisomerization; and demonstrate the possibility of attaining the same limit as SQ for conversion of
  solar energy to stored chemical energy in an STF device. (By contrast, previous analyses showed significantly lower limits.)

 Previous to the SQ work,
 researchers had found PV efficiency limits based on empirical models, which could only demonstrate where the current approaches
 to silicon solar cells might lead, but could not show the potential of other ideas that had not yet been considered.
 Understanding the SQ limit suggested the benefit of new strategies for photovoltaics such as spectrum splitting,
 multi-junctions, intermediate bands, hot carriers, multiple exciton generation, singlet fission, etc. \cite{PolmanAtwater}, or
 hybrid devices using conversion to heat as well as electricity \cite{Branz}.
 Similarly this analysis can inspire new paradigms for STFs -- indeed, upconversion \cite{Borjesson_JMCA} and hybrid solar thermal
 devices \cite{Dreos} have already been examined in the context of STFs -- and point the way to overcoming the limits we show here.

\begin{figure}[t]
\includegraphics[scale=0.60]{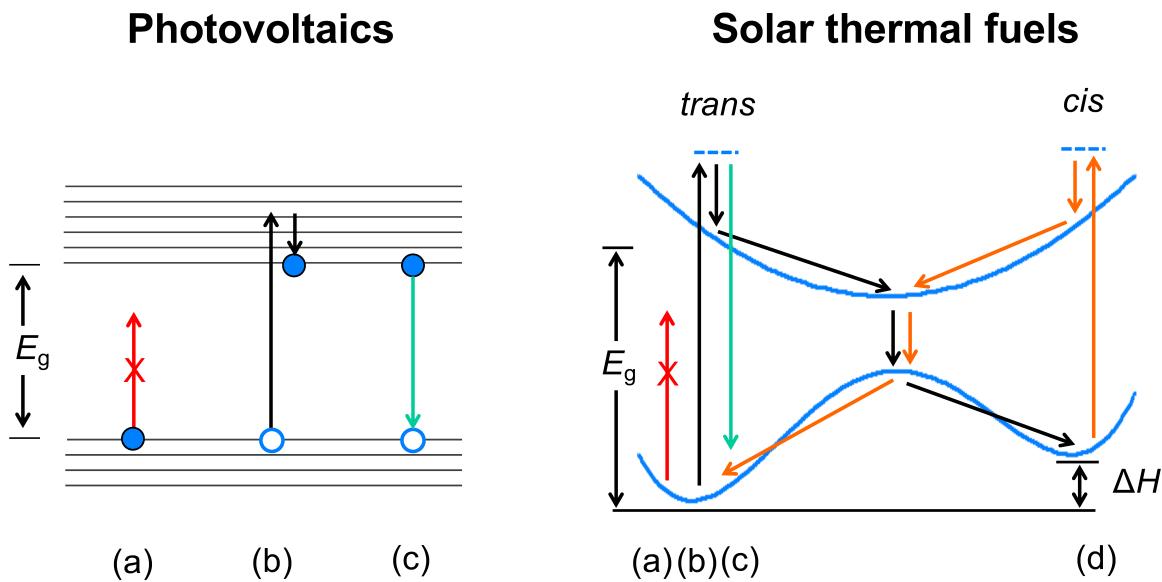}
\caption{\label{fig:cf_pv}
Comparison of basic processes in a band diagram for photovoltaics and a potential-energy surface for solar thermal fuels.
\textit{Photovoltaics}: (a) Photons with energy below the gap $E_g$ are not absorbed.
(b) Photons with energy above the gap are absorbed. The resulting carriers thermalize to the band edge and then have energy $E_g$.
(c) Radiative recombination of the excited carriers is a loss mechanism.
\textit{Solar thermal fuels}: (a) Photons with energy below the gap $E_g$ are not absorbed.
(b) Photons with energy above the gap are absorbed by \textit{trans}. The molecule relaxes to the lowest excited state at the \textit{trans} geometry,
relaxes on the potential-energy surface of that excited state, drops to the ground state, and further relaxes in the ground state to the
\textit{cis} geometry, storing an energy $\Delta H$.
(c) Fluorescence from the excited state (quantum yield $< 1$) is a loss mechanism.
(d) The reverse photoisomerization process -- absorption by \textit{cis} and conversion to \textit{trans} --
undoes the energy storage process and is another loss mechanism.
}
\end{figure}

\begin{table}[t]
\begin{ruledtabular}
  \begin{tabular}{ll}
    Photovoltaics & Solar Thermal Fuels \\ \hline\hline
    electrical power & energy storage \\
    current & conversion rate \\
    voltage & chemical potential difference \\
    short-circuit condition & thermal equilibrium \\
    open-circuit condition & photostationary state \\
    radiative recombination & fluorescence \\
    non-radiative recombination & unproductive relaxation \\
  \end{tabular}
\end{ruledtabular}  
  \caption{\label{table:processes}
    Comparison of parallel concepts between photovoltaics and solar thermal fuels.
    Key differences are the possibility of significant depletion of the ground state in STFs but not PV, the new concept
    of reverse photoisomerization in STFs, and the fact that the independent variable is {\it cis} fraction not the voltage.
    }
\end{table}

We begin by reviewing the SQ analysis and showing the analogy between PV and STFs. The basic processes are diagrammed in Figure \ref{fig:cf_pv}.
The SQ limit considers that each photon incident on the cell is not absorbed if it is below the band gap (``below-gap losses''); if it is above the
band gap, it is absorbed, but the resulting electron and hole quickly relax to the band edges and provide only energy equal to the band gap
(``above-gap losses''). These two loss mechanisms are the most important, and certainly apply to STFs. Consider the schematic potential-energy
surfaces for azobenzene. Initially light must have energy of at least $E_g$ to be absorbed by {\it trans}, and then quickly loses any excess energy beyond
that, as in a solar cell. However, after that further losses occur: the excitation relaxes on the excited-state surface to the minimum. De-excitation
to the ground state causes a further loss, as does relaxation on the ground-state surface to {\it cis}, at an enthalpy $\Delta H$ above {\it trans}.
We note that a distinction between absorption threshold and useful energy is in common
with systems that relax to a dark state, such as an indirect gap in a semiconductor or a triplet molecular state.

The simple model above does not take into account two other important loss mechanisms considered by SQ: radiative recombination, and voltage loss.
While non-radiative recombination might be reducible to zero, radiative recombination is absolutely required by detailed balance: if the cell can absorb,
it can emit. Thermally excited electron-hole pairs, populated according to the Boltzmann distribution at 300 K, can recombine and emit photons. Moreover,
the population is dependent on the voltage, thus defining the $I$-$V$ characteristics of the cell, as a maximum-power point has to be found between
the extremes of open circuit with maximum voltage but no current, and short circuit with no voltage and maximum current.
The voltage loss is the difference between the open-circuit voltage and the voltage at maximum power.

Since STFs are not electrical devices, these considerations may seem unrelated, but in fact the analogy with photovoltaics can be carried quite far.
Corresponding concepts are compared in Table \ref{table:processes}.
To begin, consider the Gibbs free energy $G = H - TS$ of a solution of an STF molecule. For concreteness, we will refer to the stable isomer as {\it trans} and the
higher-energy metastable isomer as {\it cis}, as for azobenzene, but the analysis is general.
The Gibbs free energy is the relevant thermodynamic quantity for
determining the heat released in a system at constant pressure and temperature \cite{McQuarrie_Simon}, as in the STF discharge,
and its sign determines whether a process is
spontaneous or not. Previous STF works have analyzed only the enthalpy $H$, thus working in some sense in a $T \rightarrow 0$ limit.

Let the fraction of molecules which are in the {\it cis} isomer be $x$ and in the {\it trans} isomer be $1 - x$.
(We assume a dilute solution to ensure ``ideal solution'' behavior; at high concentrations or with strong interactions between solute molecules, different
  equations than those below, with more parameters, may be required, such as the ``regular solution'' model \cite{McQuarrie_Simon}.)
Then thermal equilibrium in the dark will satisfy
\begin{eqnarray}
\frac{x}{1 - x} = K = e^{- \Delta G^{0} / k T}
\end{eqnarray}
where $K$ is the equilibrium constant, $\Delta G^{0}$ is the difference in Gibbs free energy per molecule between {\it cis} and {\it trans} under standard
conditions, $k$ is the Boltzmann constant, and $T$ is the temperature of the solution. The Gibbs free energy will vary as a function of the ratio between
{\it cis}/{\it trans} fractions $Q = x/(1-x)$, according to
\begin{eqnarray}
\Delta G \left( Q \right) = \Delta G^{0} + k T {\rm\ ln \ } Q
\end{eqnarray}
In equilibrium, $Q$ = $K$, and then
\begin{eqnarray}
\Delta G \left( K \right) = \Delta G^{0} + k T {\rm\ ln \ } K  = 0
\end{eqnarray}
From this equation, we can observe that an STF solution in equilibrium irradiated with sunlight has initial energy storage rate of zero, since $\Delta G = 0$,
even though the rate of conversion of molecules is maximum. This condition is thus analogous to the short-circuit condition for photovoltaics, since $\Delta G$
corresponds to voltage and conversion rate to current. As $Q$ increases due to the incident light, $\Delta G$ too will grow. This important effect was
not considered in previous analyses \cite{Bren,Kucharski_NatChem,Borjesson}.
We can integrate to find the total free energy stored, when cycling between two compositions $x_1$ and $x_2$:
\begin{eqnarray}
  \label{eq:storage}
  \Delta G_{\rm tot} = \int_{x_1}^{x_2} \left[ \Delta G^{0} + k T {\rm\ ln \ } \frac{x}{1-x} \right] dx
  = \left[ \Delta G^{0} x + kT x \ln x + kT \left(1 - x\right) \ln \left( 1 - x\right) \right]_{x_1}^{x_2}
\end{eqnarray}
a familiar expression from entropy of mixing, depending on temperature and fraction $x$ as well as the intrinsic molecular quantity $\Delta G^{0}$.

The rate of conversion of molecules from {\it cis} to {\it trans}, given rate constants $k_{\rm c}$ and $k_{\rm t}$ under the given illumination conditions, is
\begin{eqnarray}
\frac{dx}{dt} = \left( 1 - x \right) k_{\rm t} - x k_{\rm c}
\end{eqnarray}
where
\begin{eqnarray}
k_{\rm t} = \int I \left( \omega \right) \sigma_{\rm t} \left( \omega \right) \phi_{{\rm t} \rightarrow {\rm c}} \left( \omega \right) d\omega
\end{eqnarray}
and similarly for {\it cis}. $I$ is the incident solar photon flux (photons per time per area), which we approximate as the blackbody spectrum at
6000 K, as in the SQ analysis.
$\sigma_{\rm t}$ is the absorption cross-section, and $\phi_{{\rm t} \rightarrow {\rm c}} $ is the photoisomerization quantum yield from {\it trans} to {\it cis}.
Thus the conversion rate declines over time as $1 - x$ falls and $x$ grows. Eventually a new equilibrium in the presence of the light is established,
called the ``photostationary state'' \cite{Bandara} in which $dx/dt = 0$, in which case the ratio of fractions must be
\begin{eqnarray}
Q_{\rm max} = \left. \frac{x}{1 - x} \right|_{\rm max} = \frac{k_{\rm t}}{k_{\rm c}}
\end{eqnarray}
This ratio represents a maximum in the sense that continued irradiation will not result in further conversion of {\it trans} to {\it cis}. In fact, if the ratio
were higher, incident light would actually promote a net conversion the other way, towards the photostationary state. This condition is analogous to open-circuit
condition for photovoltaics, since $\Delta G$ is maximum but the conversion rate is zero. The composition of the photostationary state is key for the
stored energy density, representing the maximum $x_2$ possible in Equation \ref{eq:storage} and is an important target for STF design. %

The calculation of the constants $k_{\it c}$ and $k_{\it t}$ is complicated: while the absorption cross-section is straightforward,
the quantum yield is difficult to measure experimentally,
and challenging to obtain theoretically, involving calculation of non-adiabatic excited-state dynamics after light absorption \cite{Neukirch}.
The quantum yield depends sensitively on solvent and excitation energy \cite{Bandara}, and on functionalization, which may cause sensitization, quenching,
or modification of potential-energy surfaces \cite{Ceroni, Bren}.
Adsorption on a metal surface  \cite{Comstock, Comstock2} or packed templating on carbon nanotubes \cite{Kucharski_NatChem}
can dramatically reduce quantum yields, showing a key role of the environment.

However, we can put a simple limit on the photostationary state ratio, $Q_{\rm max}$, from energy conservation. Incident photons
must have at least a threshold energy $E_g$ in order to be absorbed by ${\it trans}$. Therefore, $\Delta G$ cannot exceed this value:
\begin{eqnarray}
E_{\rm g} \ge \Delta G \left( Q_{\rm max} \right) = \Delta G^{0} + k T {\rm\ ln \ } Q_{\rm max}
\end{eqnarray}
The resulting constraint on the {\it cis} fraction in the photostationary state is
\begin{eqnarray}
x \le \frac{1}{1+e^{-\left( E_{\rm g} - \Delta G^{0} \right) / k T}}
\end{eqnarray}

The difference between $E_g$ and $\Delta H$ appears as a loss in the potential-energy surface, due to contributions including the
barrier in the ground state $\Delta H^{\ddagger}$, and was considered as a fundamental constraint in the work of B\"{o}rjesson \textit{et al.}.
However, considering an ensemble at finite temperature, this need not be the case. The population $x$ of products can build up, increasing
their free energy, up to the limit just cited, $E_{\rm g} \ge \Delta G$, irrespective of $\Delta H$. Considering specifically the barrier height,
we note that transition-state theory \cite{McQuarrie_Simon} for thermal reversion assumes that the molecules at the barrier are in thermal equilibrium
with those in the metastable state, \textit{i.e.} no free-energy difference between the top of the barrier and the product \textit{cis} molecules.
As a result, the barrier height $\Delta H^{\ddagger}$ does not necessarily imply any loss of free energy, and need not be considered in our
efficiency analysis, although of course it is critical for the storage lifetime \cite{Kolpak_NanoLett}.

We have identified conditions analogous to open circuit and short circuit in photovoltaics. We can continue with an analogy to the $I$-$V$
characteristics of photovoltaics. For STF, this plot is of conversion rate of molecules \textit{vs.}
free-energy difference, with the different points on the curve
corresponding to different values of $x$. The actual rate of energy storage, like $P = I V$ in an electrical device, is
\begin{eqnarray}
P_{\rm storage} = \frac{dx}{dt} \Delta G
\end{eqnarray}
We can find the ``maximum power point'' ($x$ that maximizes $P_{\rm storage}$) by solving $dP_{\rm storage}/dx = 0$.

The efficiency $\eta$ is given by
\begin{eqnarray}
  \eta \left( x \right) = \frac{P_{\rm storage}}{P_{\rm incident}} =
  \frac{\left[ \left(1 - x\right)k_t - x k_c \right] \left[\Delta G^0 + kT \ln \frac{x}{1-x}\right]}
       {A_{\rm mol} \int \hbar \omega I \left( \omega \right) d \omega}
\end{eqnarray}
where $A_{\rm mol}$ is an effective molecular area (which will cancel out in the final result).
This equation is not a limit but an actual efficiency (assuming only independent molecules in an ideal solution) which can be computed
if the properties involved are known.

Now we will consider bounds on the rate constants $k_{t}$ and $k_{c}$, depending on the photoisomerization quantum yield.
We can put a simple bound on the quantum yield via consideration of fluorescence from molecules in the excited state,
which is analogous to radiative electron-hole recombination in PV. An excited molecule may relax (radiatively or non-radiatively)
to the ground state at any point along its path from {\it trans} to {\it cis}; at some points this relaxation will produce {\it trans} and
at others will produce {\it cis}. What we can say for certain is that the vibronic states reached by initial excitation from the
{\it trans} ground state can fluoresce and relax back to a {\it trans} structure, and so this process sets an upper bound on the quantum yield.

Let $B_{t}$ be the rate constant for absorption by {\it trans}, the same as $k_{t}$ if the quantum yield were unity.
\begin{eqnarray}
B_t = \int I \left( \omega \right) \sigma_{t} \left( \omega \right) d\omega
\end{eqnarray}
Then the absorption rate is $B_t x_t$. According to SQ's analysis and detailed balance,
a similar quantity will govern radiative recombination
back to {\it trans}, with the modifications:
\begin{enumerate}
\item the solar photon flux is replaced by the blackbody spectral intensity at room temperature ($T$ = 300 K),
\begin{eqnarray}
  I_{\rm bb} \left( \omega \right) = \frac{2 \omega^2}{\pi c^2} \frac{1}{e^{\hbar \omega/kT} - 1},
\end{eqnarray}
\item there is an additional factor of 2 to account for the fact that the device can only absorb from one illuminated side
but can radiate from both sides, and
\item the emission probability is given by $\sigma_t$ multiplied by the Boltzmann factor $e^{E_g / kT}$
(using the energy difference between the ground and excited states of {\it trans}),
since the emission is proportional to the occupation of excited states, which are increased by this factor when the system
is driven out of equilibrium under illumination.
\end{enumerate}
This radiative recombination coefficient is
\begin{eqnarray}
A_t = 2 e^{E_g / kT} \int I_{\rm bb} \left( \omega \right) \sigma_{t} \left( \omega \right) d\omega
\end{eqnarray}
The emission rate then is $A_t x_s$, where $x_s$ is the fraction of molecules in the excited state.

An upper bound on the conversion rate to {\it cis} comes from taking this radiative recombination as the only process preventing an excited {\it trans}
molecules from converting to {\it cis}:
\begin{eqnarray}
\frac{dx}{dt} \le B_{t} \left( 1 - x \right) - A_{t} x_s - x k_{c}
\end{eqnarray}
The first two terms represent the rate due to absorption by {\it trans}, $k_t x_t$.
\begin{eqnarray}
  k_t x_t \le B_t x_t - A_t x_s = \int \left( I \left( \omega \right) x_t - 2 I_{\rm bb} \left( \omega \right) e^{E_g / kT} x_s \right)
  \sigma_{t} \left( \omega \right) d\omega
\end{eqnarray}
Comparing to the expression for $k_t$,
we find in fact a bound on the quantum yield of photoisomerization across the spectrum:
\begin{eqnarray}
\phi_t \left( \omega \right) \le 1 - \frac{2 I_{\rm bb} \left( \omega \right)}{I \left( \omega \right)} e^{\Delta G_{st} / kT}
\end{eqnarray}
This is expressed in terms of the free-energy difference between the ground and excited states of {\it trans},
which is not a quantity that is easily measured or controlled.
Instead, we can use the inequality $\Delta G_{st} \ge \Delta G$, which is required for the excited state to be able to drive the structural change to {\it cis}.
Then
\begin{eqnarray}
\phi_t \left( \omega \right) \le 1 - \frac{2 I_{\rm bb} \left( \omega \right)}{I \left( \omega \right)} e^{\Delta G / kT}
\end{eqnarray}
This quantum yield bound decreases as a function of conversion percentage (through $e^{\Delta G / kT} = Q e^{\Delta G^{0} / kT}$),
and therefore makes a contribution to the $I-V$ characteristics of the STF.
Moreover, we have shown that the quantum yield cannot reach unity even in principle, due to microscopic reversibility.
This bound can be used in place of the simple assumption of $\phi_t = 1$ in previous efficiency analyses. %

Following the SQ approach, we can let the absorption probability for {\it trans} be 1 above the band gap and 0 below, which can be approached in practice
by making the device thick enough so that all incoming light is absorbed. This is the maximum possible absorption, which will lead to the best efficiency,
and implies a cross-section equal to $A_{\rm mol}$.
\begin{eqnarray}
\sigma_t \left( \omega \right) = \left\{ \begin{array}{lr}
       0 : \hbar \omega < E_g \\
       A_{\rm mol} : \hbar \omega > E_g
     \end{array}
\right.
\end{eqnarray}
On the other hand, absorption by {\it cis} reduces the efficiency, and so we will take $\sigma_c \left( \omega \right) = 0$, the lowest possible
absorption. (While quantum-mechanical sum rules require some absorption, it can be pushed arbitrarily far out of the solar spectrum to achieve
a similar result.) This limit also sets the {\it cis} $\rightarrow$ {\it trans} reverse photoisomerization to zero, removing this loss from
consideration. That could also happen via $\phi_{c \rightarrow t} = 0$, as for the ``one-way'' photoisomerizable molecules such as
dihydroazulene/vinylheptafulvene which do not exhibit a reverse process \cite{Kucharski_review}.

\begin{figure}[t]
\includegraphics[scale=0.6]{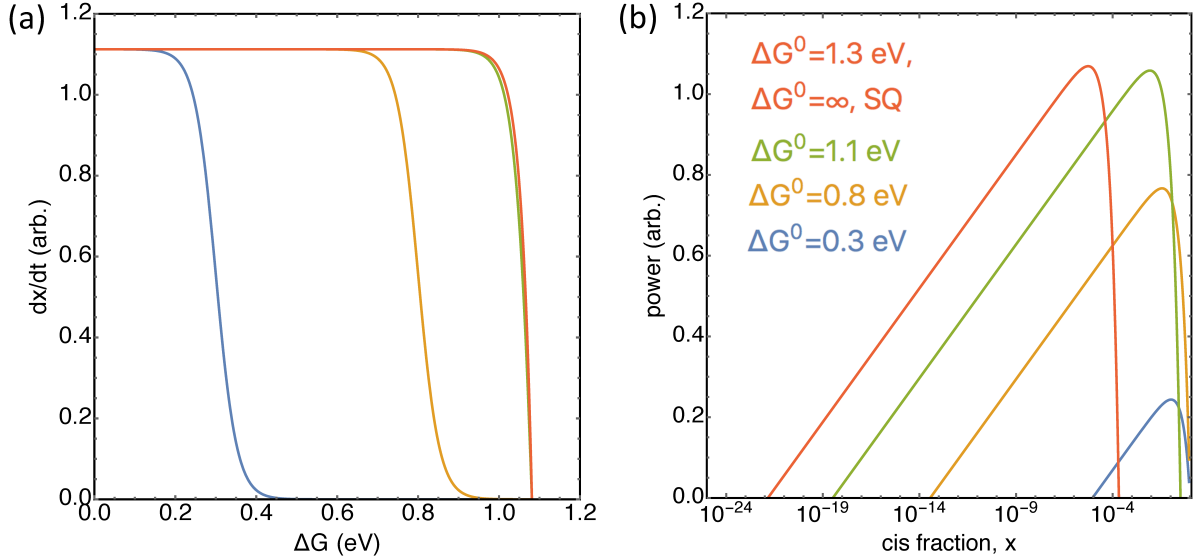}
\caption{\label{fig:i_v_power}
  (a) Conversion rate vs free-energy difference, a curve analogous to $I$-$V$ characteristics, but traced out by varying the \textit{cis} fraction. The lines for
  $\Delta G^0 = \infty$, $\Delta G^0 = $ 1.3 eV, and the Shockley-Queisser $I-V$ characteristic shape are indistinguishable.
(b) The power being stored as a function of \textit{cis} fraction. The legend for $\Delta G^0$ applies to both plots, and $E_g$ = 1.3 eV.
}
\end{figure}

We have now a simplified model giving an upper bound to the efficiency, involving as parameters only $E_g$ and $\Delta G^{0}$, both of which
can be straightforwardly measured and computed theoretically:
\begin{eqnarray}
  \eta \left( x \right) =
  \frac{\left[ \left(1 - x\right) \int_{E_g} I \left( \omega \right) d\omega - 2 x e^{\Delta G^{0}/kT} \int_{E_g} I_{\rm bb} \left( \omega \right) d\omega \right] \left[\Delta G^0 + kT \ln \frac{x}{1-x}\right]}
       {\int \hbar \omega I \left( \omega \right) d \omega}
\end{eqnarray}
The numerical solutions of the conversion rate \textit{vs.} free-energy difference, a curve analogous to $I$-$V$ characteristics,
are shown in Figure \ref{fig:i_v_power} for the case $E_g$ = 1.3 eV and various values of $\Delta G^{0}$. Changing $\Delta G^{0}$ has little effect
on the conversion rate at fixed $\Delta G$, but it strongly affects the maximum $\Delta G$ attainable (at the photostationary state). This maximum
$\Delta G$ increases with $\Delta G^{0}$ but saturates at 1.1 eV and the curves become indistinguishable beyond that.
The power is also plotted as a function of $x$, which shows a slow rise and steep fall. The smallest $x$ where power generation occurs
is the thermal population, which of course decreases with increasing $\Delta G^{0}$. The value of $x$ at which the maximum power is attained
falls with increasing $\Delta G^0$, showing a trade-off between maximum rate of energy storage and the maximum amount of energy that can be stored
(as in equation \ref{eq:storage}).

In photovoltaics, power electronics can be used to vary the resistive load across the device in order to operate close to the maximum power point.
In STF, we need to control the {\it cis} fraction to do the equivalent. For example, the rate at which the solution flows through a plate where it is exposed to
the sun \cite{poulsen_EES} can be optimized (given the charging rate) in order to keep the solution near the maximum power point, if one wished to achieve the
maximum energy storage rate. Of course, doing so would result in quite a small conversion percentage and thus not be the best choice for energy storage density.
An alternate possibility is controlling {\it cis} fraction via differential solubility or density of the two isomers in a liquid phase.

\begin{figure}[t]
\includegraphics[scale=0.6]{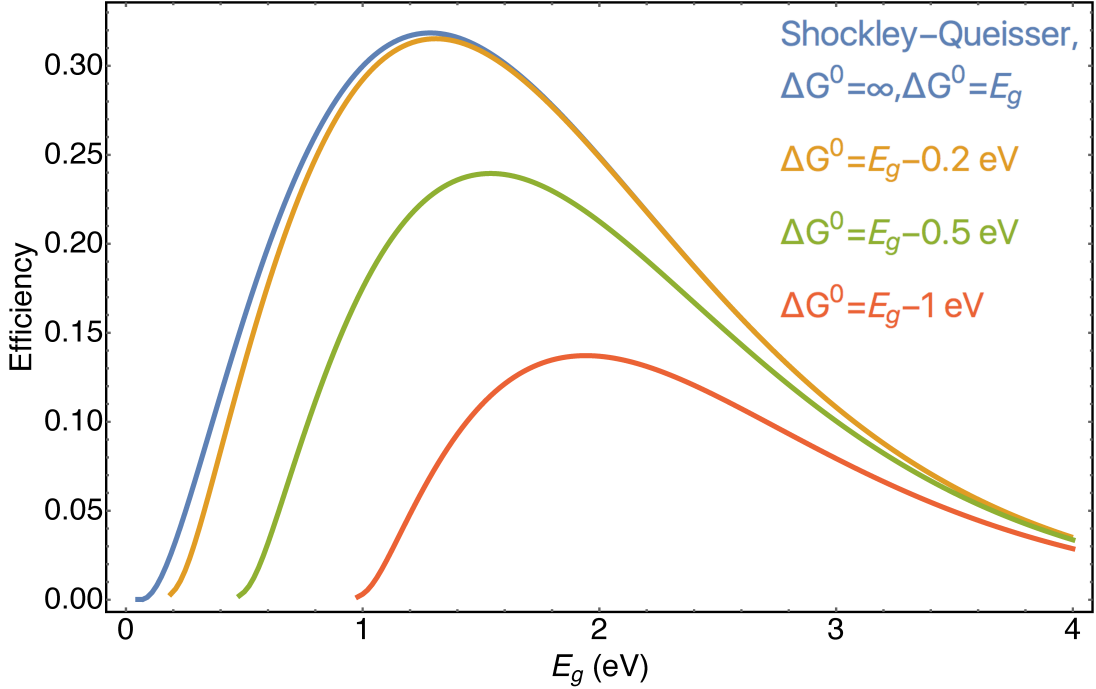}
  \caption{\label{fig:efficiency}
  Efficiency limit in converting incident solar energy to stored chemical energy, as a function of band gap,
  at the optimal {\it cis} fraction $x$ for each gap. The ultimate limit is 32\% at $E_g = 1.27$ eV, in the limit $\Delta G^0 \rightarrow \infty$
  which is identical to the Shockley-Queisser limit. The curve for $\Delta G^0 = E_g$ is almost indistinguishable from these limits, but as $\Delta G^0$
  is reduced, the maximum efficiency drops and moves to a higher value of $E_g$. The point at which efficiency rises above zero is approximately $\Delta G^{0}$.
}
\end{figure}

Consider the case $\Delta G^{0} \rightarrow \infty$. This is consistent with the requirement that free energy decreases,
which stipulates only $\Delta G < E_g$ (as we used for the limit on the photostationary state). In this limit, for a given $\Delta G$, $x$ goes to 0,
removing the loss of absorption due to depletion of the \textit{trans} molecules, and remarkably reducing the efficiency equation to one equivalent to SQ
(following the translation of concepts in Table \ref{table:processes}):
\begin{eqnarray}
\eta_{\rm \Delta G^{0} \rightarrow \infty} \left( \Delta G \right) =
  \frac{\left[ \int_{E_g} I \left( \omega \right) d\omega - 2 e^{\Delta G/kT} \int_{E_g} I_{\rm bb} \left( \omega \right) d\omega \right] \Delta G }
       {\int \hbar \omega I \left( \omega \right) d \omega}
\end{eqnarray}
For lesser values of $\Delta G^{0}$, the efficiency is reduced due to {\it trans} depletion, but $\Delta G^{0} \ge E_g$ is sufficient to obtain almost
the maximum efficiencies.
These results are plotted in Figure \ref{fig:efficiency}, exhibiting the maximum of 32\% at 1.27 eV for $\Delta G^{0} \rightarrow \infty$.
Comparing to the experimentally estimated efficiency of 0.07\% for the Ru-dithiafulvalene system \cite{poulsen_EES}, it is clear there is the possibility
of great improvement in STFs.

\begin{figure}[t]
\includegraphics[scale=0.7]{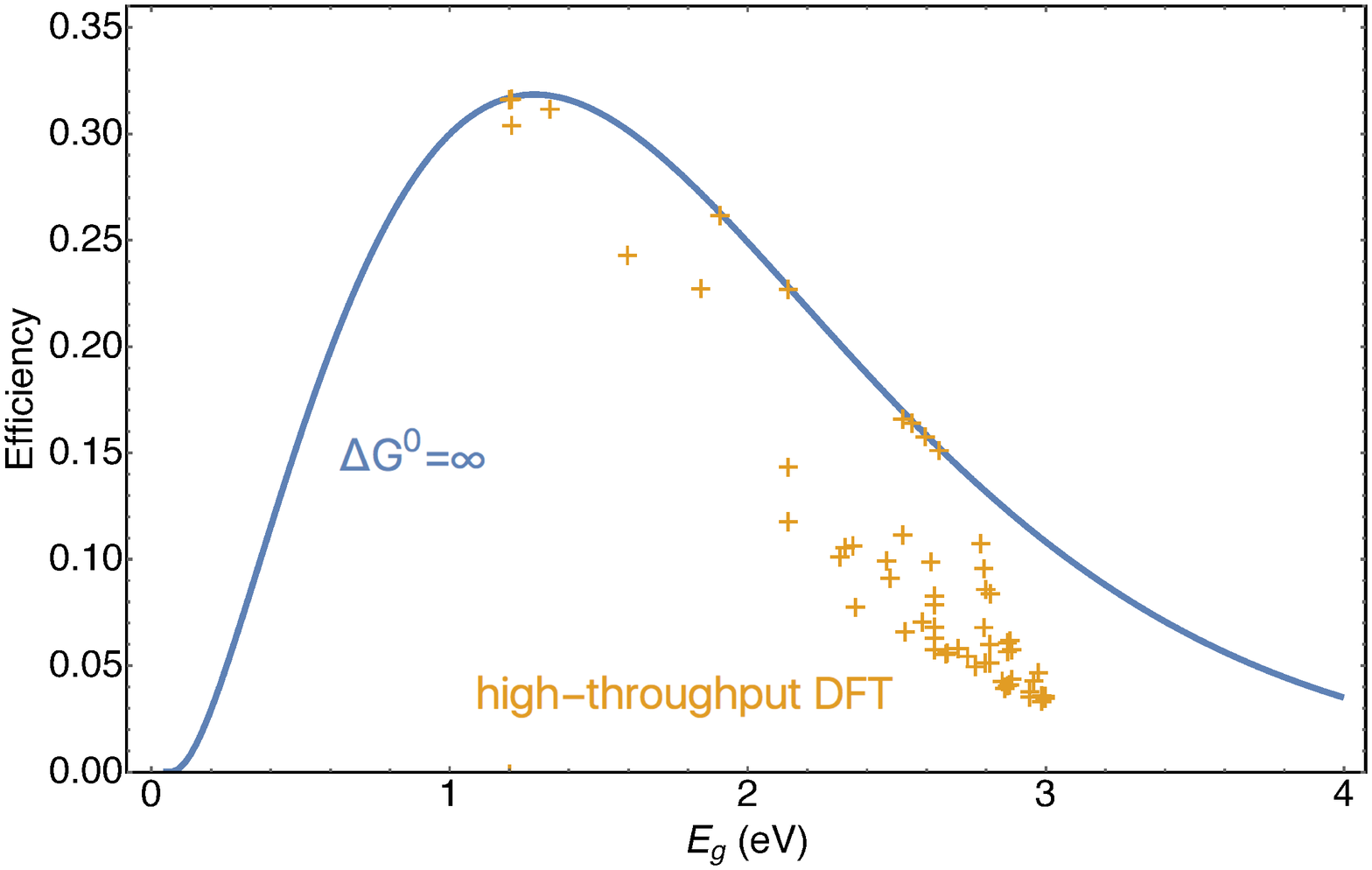}\\
\caption{\label{fig:highthroughput}
  Efficiency limits for converting incident solar energy to stored chemical energy, for 62 candidate azobenzene derivative molecules
  identified in a high-throughput search using density-functional theory \cite{Liu}. The optical gap $E_g$ is the Kohn-Sham band gap + 0.9 eV, and
  $\Delta G^{0}$ is the total energy difference. The curve shows the limit $\Delta G^{0} \rightarrow \infty$.
}
\end{figure}

High-throughput screening of molecules for STF applications has already begun, using azobenzene derivatives \cite{Liu} and later norbornadiene/quadricyclane
derivatives \cite{Kuisma}. These works have assessed their candidates only by considering parameters separately, or with regard to the older attempt at an
efficiency limit \cite{Borjesson}. Our
improved and more fundamental limit enables a more powerful screening without unnecessary assumptions. We reassess the azobenzene derivatives of \cite{Liu}
in Figure \ref{fig:highthroughput}, using the estimated $E_g$ (as PBE Kohn-Sham gap + 0.9 eV, as in that work) and $\Delta G^{0}$ as the total energy difference
(neglecting effects of vibrational entropy or volume changes).
We observe that several molecules indeed come very close to the maximum efficiency limit. Many of the molecules have a large enough $\Delta G^0$ to reach the maximum efficiency
for their $E_g$, but their potential efficiency is limited by having too large $E_g$. Therefore, smaller $E_g$ should be a design goal to find improved STF molecules.

We underscore numerous important new aspects in our approach to understanding STF efficiency limits.
We do not assume a quantum yield $\phi = 1$
but in fact derived an upper bound for $\phi$ dependent on the extent of charging. We showed that the ground-state barrier to thermal reversion, $\Delta G^{\ddagger}$,
does not inherently cause a loss and does not enter into our final limit. We demonstrated the critical importance of the free energy, not just the enthalpy,
because of the effect of temperature and entropy of mixing. Most importantly, we found that thermodynamic considerations provide the same 32\% efficiency limit,
as same optimal band gap, as in the Shockley-Queisser analysis.

This analysis of the fundamental limits to STF efficiency helps to benchmark candidate materials and devices
against potential performance, identify the weak points that are most important to improve, and focus thinking on applications by showing what is the best
performance we can expect.
Our results demonstrate that STFs may match the peak efficiencies of PV (despite the much lower performance of current STF devices),
although further losses may occur in conversion of the stored chemical energy, into {\it e.g.} electricity. This level of performance
is very promising for applications where heat will be used directly \cite{HEATS}, and also where the storage feature is particularly valued.
We believe this thermodynamic approach to STF efficiency could enable the development of new STF materials and paradigms,
and provides new insights into photochemistry generally, for example by demonstrating
the existence of limits to quantum yield regardless of details of reaction pathways.

\section{acknowledgments}

We would like to acknowledge helpful discussions with Alexie Kolpak, Yun Liu, Tim Kucharski, and Jee Soo Yoo.
This work was supported by the U. S. Advanced Research Projects Agency - Energy under Grant DE-AR0000180.


\end{document}